\documentclass[]{nature}
\usepackage{times}
\usepackage[super,numbers]{natbib}
\usepackage[margin=1in]{geometry}                         
\usepackage{amsmath}
\usepackage{pdflscape}
\usepackage{amsfonts}
\usepackage{amssymb}
\usepackage{graphicx}
\usepackage[normalem]{ulem}
\usepackage{hyperref}
\usepackage{float}
\usepackage{subcaption} 
\usepackage{authblk}
\usepackage[left]{lineno}

\makeatletter
\let\saved@includegraphics\includegraphics
\AtBeginDocument{\let\includegraphics\saved@includegraphics}
\renewenvironment*{figure}{\@float{figure}}{\end@float}
\makeatother

\title{Persistent plasma waves in interstellar space detected by Voyager 1} 
\date{}
\author[1]{Stella Koch Ocker\footnote{corresponding author, sko36@cornell.edu}}
\author[1]{James M. Cordes}
\author[1]{Shami Chatterjee}
\author[2]{Donald A. Gurnett}
\author[2]{William S. Kurth}
\author[2]{Steven R. Spangler}
\affil[1]{Department of Astronomy and Cornell Center for Astrophysics and Planetary Science, Cornell University, Ithaca, NY 14850, USA}
\affil[2]{Department of Physics and Astronomy, University of Iowa, Iowa City, IA 52242, USA}

\begin{document}

\maketitle

In 2012, Voyager 1 became the first in situ probe of the very local interstellar medium \citep{2013Sci...341.1489G}. The Voyager 1 Plasma Wave System has given point estimates of the plasma density spanning about 30 au of interstellar space, revealing a large-scale density gradient \citep{2019NatAs...3.1024G, 2020ApJ...900L...1K} and turbulence \citep{2019NatAs...3..154L} outside the heliopause. Previous studies of the plasma density relied on the detection of discrete plasma oscillation events triggered ahead of shocks propagating outwards from the Sun and used to infer the plasma frequency and hence density \citep{2015ApJ...809..121G, 2021AJ....161...11G}. We present the detection of a class of very weak, narrowband plasma wave emission in the Voyager 1 data that persists from 2017 onwards and enables the first steadily sampled measurement of the interstellar plasma density over about 10 au with an average sampling distance of 0.03 au. We find au-scale density fluctuations that trace interstellar turbulence between episodes of previously detected plasma oscillations. Possible mechanisms for the narrowband emission include thermally excited plasma oscillations and quasi-thermal noise, and could be clarified by new findings from Voyager or a future interstellar mission. The emission's persistence suggests that Voyager 1 may be able to continue tracking the interstellar plasma density in the absence of shock-generated plasma oscillation events.

\section{Main}

\begin{figure}
    \centering
    \includegraphics[width=\textwidth]{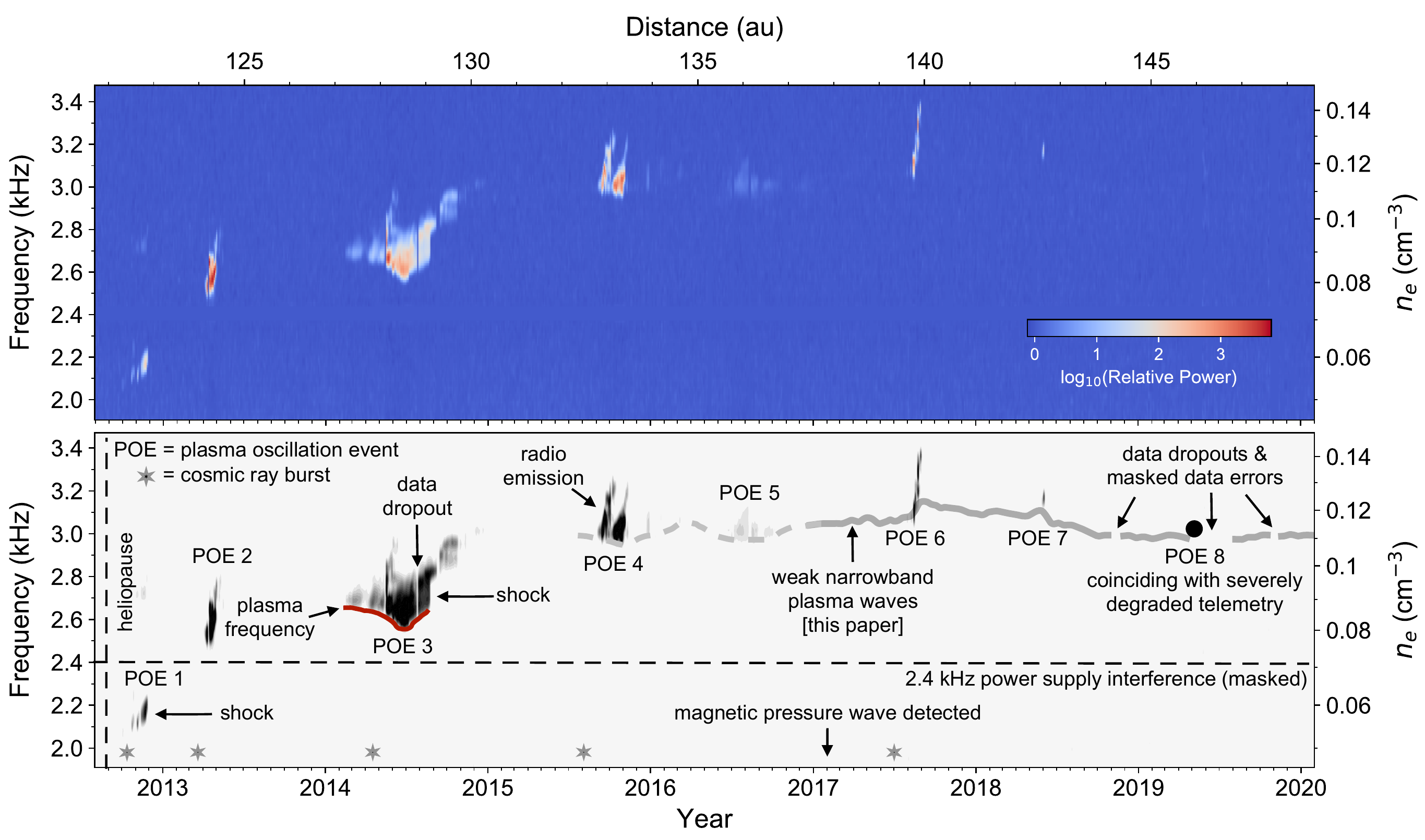}
    \caption{\textbf{Voyager 1 composite plasma wave spectrum.} Top: Frequency-time dynamic spectrum showing all of the Voyager 1 PWS wideband data available since Voyager 1 crossed the heliopause on August 25, 2012. The time resolution of the spectrum is 3 days and the frequency resolution is 0.011 kHz. Each column of pixels corresponds to a 1D spectrum that is the average of all $\approx 48$ s long observations that fall within that 3-day time bin. The individual spectra that fall within a given time bin have been equilibrated to the same noise baseline. The 2.4 kHz supply interference line is masked out, and the spectrum is smoothed with a 1D Gaussian smoothing kernel with $\sigma = 0.01$ kHz. Bottom: Schematic showing relevant features in the PWS spectrum, including the locations of previously detected plasma oscillation events (POEs). Two events have direct associations with shocks detected by the magnetometer. A magnetic pressure wave was also detected in early 2017. The lower cutoff frequency of the plasma oscillations corresponds to the local plasma frequency. The approximate times of relativistic electron bursts detected by the cosmic ray instruments are also indicated. The model of the weak, narrowband plasma wave emission presented in this paper is shown in solid gray, and the plasma frequency inferred from POE activity between 2015 and 2017 is shown in dashed gray. A POE detected in June 2019 is also shown as a black circle, but was masked in our analysis because it coincided with a period of severely degraded telemetry performance.}
    \label{fig:fullspec}
\end{figure}

\indent Since crossing the heliopause, Voyager 1 (V1) has measured the plasma density of the very local interstellar medium (VLISM) by detecting plasma oscillation events (POEs). The Plasma Wave System (PWS) measures these POEs as narrow ($\approx0.2$ to $0.4$ kHz wide) features in frequency-time spectra obtained with the wideband receiver and the plasma wave spectrometer. Fig.~\ref{fig:fullspec} is a composite dynamic spectrum showing all of the high-time resolution data available from the PWS wideband receiver since V1 crossed the heliopause until early 2020. The wideband receiver measures the voltage difference at the antenna terminals at a rate of 28,800 samples s$^{-1}$, which is converted into a frequency-time dynamic spectrum by Fourier transforming the voltage time series. The composite spectrum in Fig.~\ref{fig:fullspec} shows salient features between 1.9 and 3.5 kHz and has time and frequency resolutions of 3 days and 0.011 kHz, respectively.

\indent Eight distinct POEs are visible in the composite dynamic spectrum beginning in late 2012. While the duration of these events ranges from a couple of days to a year, they are all believed to be generated via beam-plasma instabilities in the electron foreshocks of shocks propagating out of the heliosphere \citep{2021AJ....161...11G}. The POEs share key characteristics, including: 1) a spreading in frequency associated with the excitation of higher wave modes via, e.g., Langmuir parametric decay \citep[][]{1992GeoRL..19.2187C, 1994JGR....9913363H}; 2) an upward frequency drift attributed to the increase in plasma density as the shock propagates over the spacecraft, which in two cases has been directly associated with a jump in magnetic field strength characteristic of a quasi-perpendicular shock \citep{2013ApJ...778L...3B, 2015ApJ...809..121G, 2017ApJ...843L..32K}; and 3) fast ($\sim$ sub-second level) intensity variations accompanied by smoother, broadband radio emission generated in a nonlinear mode conversion process. Most of these events are preceded by relativistic cosmic ray bursts detected by the V1 cosmic ray instruments, which indirectly provide energy estimates of $\sim 50$ eV for the electron beams generating the plasma oscillations \citep{2021AJ....161...11G}.

\begin{figure}
    \centering
    \includegraphics[width=120mm]{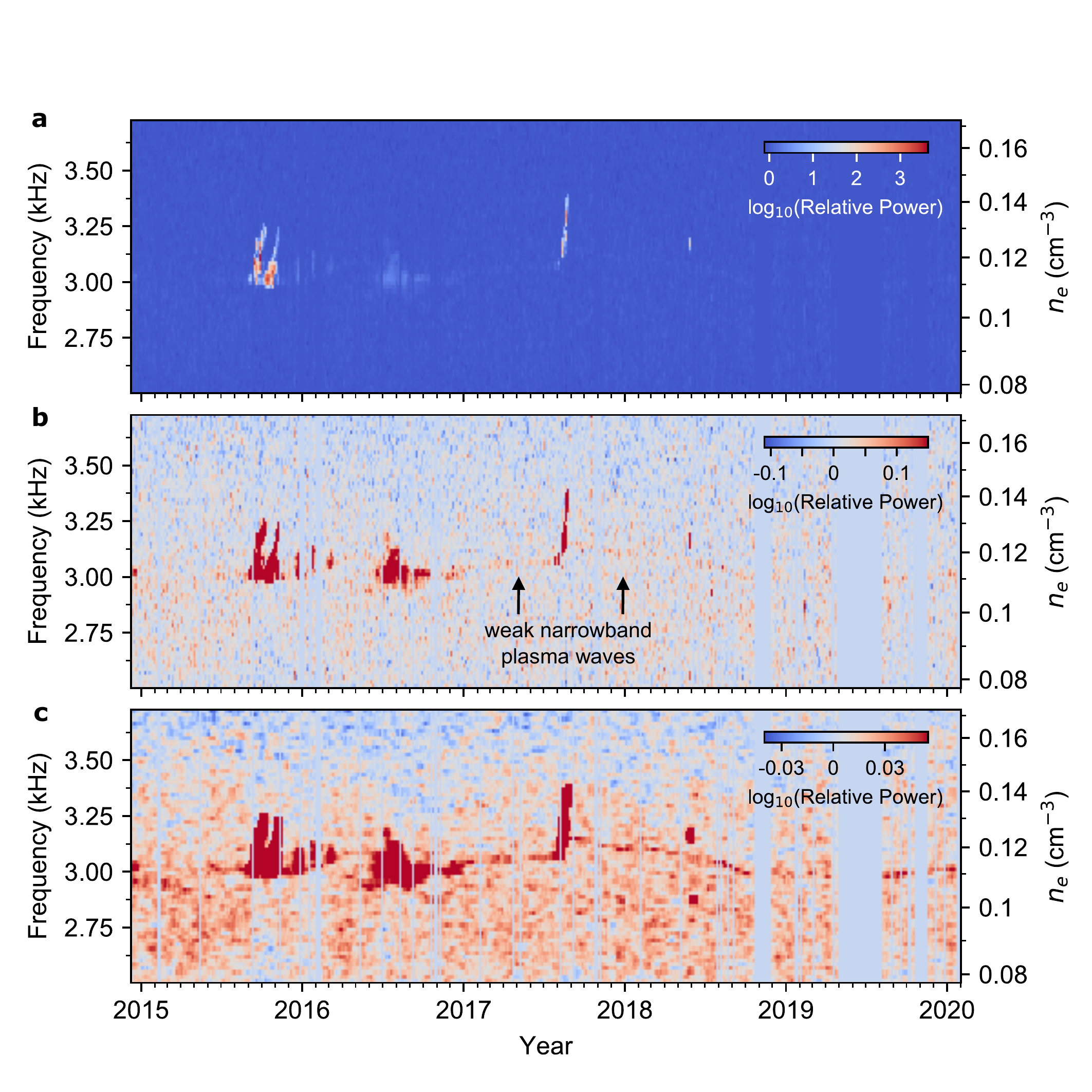}
    \caption{\textbf{Weak narrowband plasma waves in the Voyager 1 dynamic spectrum.} a) Dynamic spectrum showing the Voyager 1 PWS wideband data from December 2014 through January 2020. The spectrum was creating using the same methods as Fig.~\ref{fig:fullspec}, but with a slightly coarser frequency resolution (about 0.02 kHz) to avoid spectral ringing, and without Gaussian smoothing. Epochs contaminated by telemetry errors have been masked out; masked data and data dropouts appear as vertical stripes with zero logarithmic power. b) Same as top panel, but with a different color stretch that extends to 0.16 in logarithmic power.
    c) Same as top panel but smoothed in time using a 1D boxcar filter and a color stretch extending to a logarithmic power of 0.055.}
    \label{fig:panelspec}
\end{figure}

\indent In Fig.~\ref{fig:panelspec} we show the dynamic spectrum of the PWS wideband data from 2015 through early 2020. Fig. 2a shows the same data as in Fig.~\ref{fig:fullspec} with additional mitigation of telemetry errors. Fig. 2b shows the same dynamic spectrum but with a different color stretch that shows evidence of a new narrowband signal spanning the entire period of data available between 2017 and 2020. Fig. 2c shows the result of convolving a 1D boxcar filter in time with the dynamic spectrum. This figure also shows complicated POE activity in 2015 and 2016. The 2015 POE shows possible evidence of multiple shocks, and there is a sharp drop in the plasma frequency between March and June 2016 when V1 passes into a low-density region. The 2014 POE persists through December 2014, but it is unclear whether the spikes of intensity from early-mid 2015 in the boxcar-filtered spectrum trace the narrowband emission observed later on. 

\indent The narrowband signal shown in Fig.~\ref{fig:panelspec} constitutes a new class of plasma wave emission in the Voyager spectrum. Unlike previously detected plasma oscillations, its narrower width remains roughly constant at about $0.04$ kHz, and it persists for almost three years (barring data dropouts), corresponding to a distance traveled by the spacecraft of about 10 au. The signal-to-noise ratio (SNR) of the narrowband emission is no more than 2 in a single 48 s epoch and requires some degree of averaging to be measured (see Methods). By contrast, previously detected POEs are detectable within individual epochs at the sub-millisecond level with SNRs $>100$, and can vary by more than an order of magnitude in intensity between epochs spaced a few days apart. There is also no evidence of the narrowband signal spreading in frequency or monotonically drifting upwards in frequency. Given its narrow bandwidth, low amplitude, and years-long persistence, the narrowband plasma wave emission appears to be distinct from the shock-generated POEs.

\begin{figure}
    \centering
    \includegraphics[width=89mm]{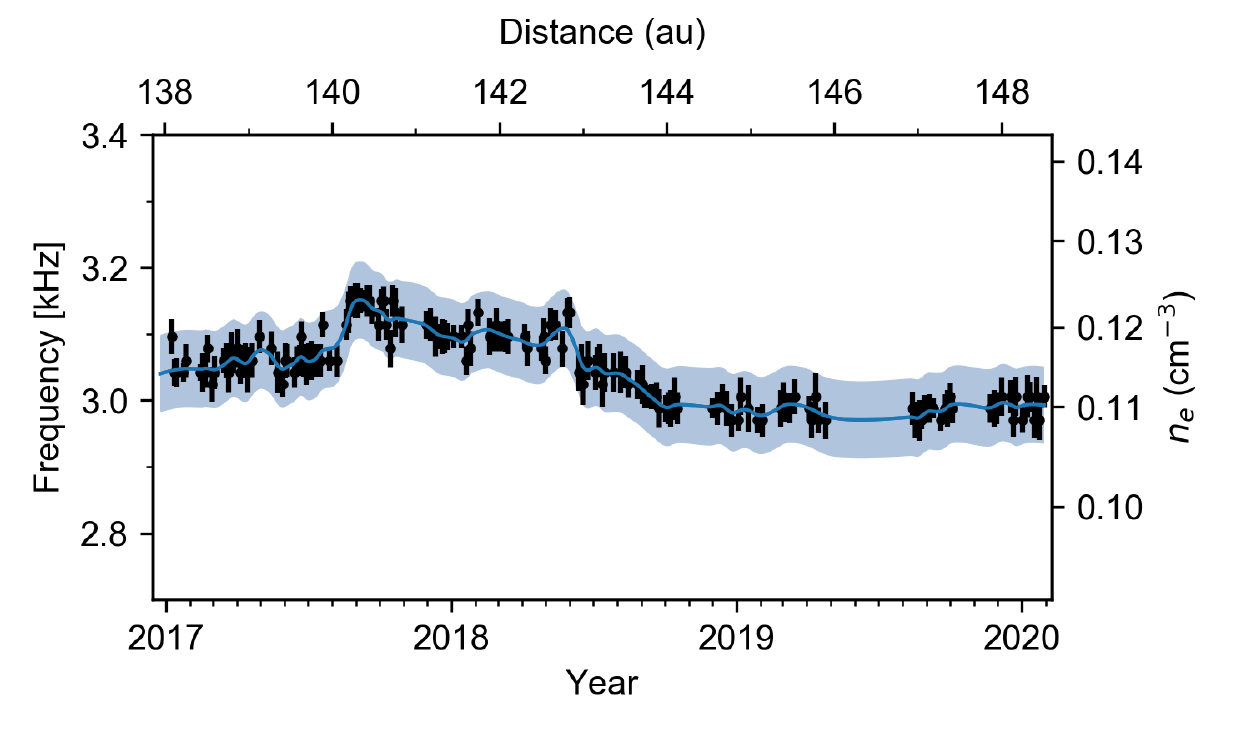}
    \caption{\textbf{Peak frequency vs. time of the narrowband plasma wave emission.} The frequencies were extracted by running a friends-of-friends algorithm on the spectrum in Fig.~\ref{fig:panelspec}b. Error bars represent the rms of the noise fluctuations in each column of Fig.~\ref{fig:panelspec}b. The blue curve and shaded region show the best-fit Gaussian process model and $95\%$ confidence intervals, respectively. The Gaussian process model is insensitive to large data gaps and treats the density as roughly constant during these gaps.}
    \label{fig:timeseries}
\end{figure}

\begin{figure}
    \centering
    \includegraphics[width=89mm]{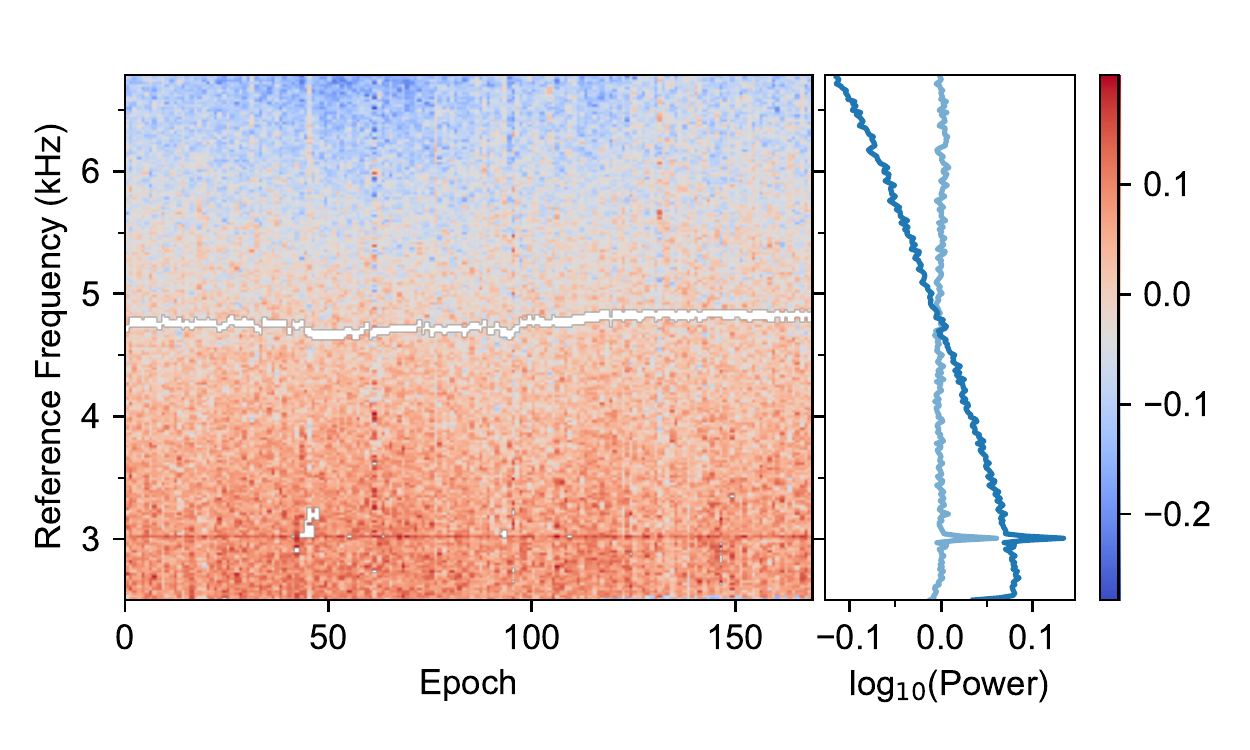}
    \caption{\textbf{De-fluctuated dynamic spectrum of the narrowband plasma wave emission.} Epochs shown are between January 2017 and February 2020 when the narrowband plasma wave emission was detected. The peak frequency of the emission was aligned to a reference frequency of 3 kHz. The 4.8 kHz harmonic of the power supply interference line was masked (white) prior to shifting the spectrum. The right-hand panel shows the de-fluctuated spectrum averaged in time with the power-law noise continuum included (dark blue) and with the continuum removed using a second-order polynomial fit to the logarithmic power vs. frequency (light blue). There is no evidence of a plasma line harmonic at 6 kHz.}
    \label{fig:shifted_ds}
\end{figure}

\indent The peak frequency of the narrowband emission vs. time is shown in Fig.~\ref{fig:timeseries}. The peak frequency was measured by running a modified friends-of-friends (FoF) algorithm \citep{1982ApJ...257..423H} on the spectrum in Fig.~2b. A Gaussian process model was fit to the frequency time series using Bayesian regression \citep{scikit-learn} to obtain a continuous model of the peak frequency vs. time, and is shown in blue in Fig.~\ref{fig:timeseries}. The dynamic spectrum was then ``de-fluctuated" by aligning the peak frequencies of the narrowband emission to a reference frequency of 3 kHz (see Fig.~\ref{fig:shifted_ds}). The de-fluctuated spectrum averaged over all epochs in which the narrowband emission was detected is also shown in Fig.~\ref{fig:shifted_ds}. Over a timespan of 70 epochs (or 210 days), the average amplitude of the narrowband emission is 0.06 (in units of log$_{10}$(power) relative to the noise baseline) and the root-mean-square (rms) of the noise is 0.005. The de-fluctuated spectrum averaged over the entire duration of the signal shows no evidence of plasma line harmonics above an rms noise threshold of 0.002. There is no evidence of the narrowband emission above an rms noise amplitude of 0.005 when the data between the POEs of 2013, 2014, and 2015 are also averaged over 210-day timespans. Given that the narrowband emission overlaps with the end of the 2016 POE (see Fig.~\ref{fig:panelspec}), it is possible that the narrowband emission is also present in 2015 and 2016. However, the low SNR of the narrowband emission makes it difficult to distinguish from the higher-intensity POEs occurring in that period. 

\indent Knowledge of the emission mechanism is required to associate the frequency of the narrowband signal with the plasma frequency ($f_p$) and infer the electron density ($n_e$). The behavior of the narrowband signal before and after the POE of September 2017 is highly suggestive that its frequency traces the plasma density. The leading edge of the 2017 POE occurs at the same frequency (3.06 kHz) as the narrowband emission prior to the POE's onset. In mid-September 2017 the plasma frequency increases to 3.15 kHz as the associated shock passes over the spacecraft. During the same period, the frequency of the narrowband emission also increases, before drifting downwards over several months after the POE. The frequency of the narrowband emission is also similar to the plasma frequencies measured from POEs in 2015 and 2016, which are consistent with $n_e\sim0.11$ cm$^{-3}$. This electron density is consistent with the mean density of the Local Interstellar Cloud estimated from carbon absorption lines in spectra of nearby stars \citep{2008ApJ...683..207R}.

\indent The weak, narrowband plasma wave emission could be generated through thermal or suprathermal processes. Thermal density fluctuations with wavelengths longer than the Debye length ($\sim 20$ m in the VLISM) can induce electrostatic oscillations at the plasma frequency, and the intensity of the fluctuations can also be enhanced by suprathermal electrons \citep{1960PhRv..120.1528S, 1965PhRv..139...55P}. Thermal plasma oscillations are routinely observed with incoherent radar scattering in Earth's ionosphere and have intensities enhanced by energetic photoelectrons \citep{1960RSPSA.259...79D, 1982JATP...44.1089C, 2017GeoRL..44.5301V}. In the VLISM, suprathermal electrons might be contributed from inside the heliosphere or near the heliopause. It's possible that an anisotropy of low-energy cosmic rays from, e.g., a nearby star could induce plasma oscillations. Any explanation of this nature must account for the emission's apparent onset about 15 au away from the heliopause and its years-long duration.

\indent An alternative explanation for the narrowband signal is quasi-thermal noise (QTN), which has been routinely observed by plasma wave instruments on Cassini, Galileo, and other Solar System missions \citep{2017JGRA..122.7925M}. QTN is generated by the quasi-thermal motions of electrons in a plasma, resulting in a local electric field. The QTN voltage spectrum peaks at the local $f_p$, and its overall shape depends on the antenna properties, with thin antenna longer than the Debye length required to mitigate shot noise \citep{2017JGRA..122.7925M}. The Voyager antenna, with an effective length of 10 meters and width of 12.5 mm, might be able to detect a marginal peak at $f_p$ in the QTN spectrum, but QTN generally predicts a much broader frequency bandwidth than we observe. A much longer antenna on a future interstellar probe would be capable of robustly detecting QTN. 

\indent The peak frequency of the narrowband emission likely tracks the plasma frequency, so we use the frequency time series in Fig.~\ref{fig:timeseries} to estimate the spectral coefficient of the electron density power spectrum. The power spectrum of electron density fluctuations in the ISM follows a roughly power-law dependence of the form:
\begin{equation}\label{eq:wavespec}
    P_{\rm \delta n_e}(\mathbf{q}) = C_{\rm n}^2 q^{-\beta}, \hspace{0.1in} q_{\rm o} \leq q \leq q_{\rm i}
\end{equation}
where $C_{\rm n}^2$ is the spectral coefficient, and $q_{\rm o}$ and $q_{\rm i}$ are the outer and inner scales of turbulence \citep{1990ARA&A..28..561R}. Scattering measurements from compact radio sources like pulsars have constrained the inner scale of turbulence in the ISM to be between about 100 and 1000 km \citep{1990ApJ...353L..29S, 2004ApJ...605..759B, 2009MNRAS.395.1391R}. A recent analysis of electron densities measured with V1 from 2012 through late 2016 showed broad consistency with a Kolmogorov power law ($\beta = 11/3$), but found a higher spectral coefficient at the smallest spatial scales $\sim$10s of meters \citep[][]{2019NatAs...3..154L, 2020ApJ...904...66L}. The magnetic field fluctuations observed by Voyager are consistent with turbulence in the VLISM being driven at an outer scale $\sim 0.01$ pc \citep{2018ApJ...854...20B}, while the inner scale may be related to the ion or electron gyroradii or the Debye length \citep{2019NatAs...3..154L}. 

\indent The densities inferred from the narrowband emission reveal variations on scales $\sim1$ au combined with discrete events on $\sim0.5$ au scales produced by shocks originating in the heliosphere. The observed variance of the density fluctuations is directly related to the density power spectrum (Eq.~\ref{eq:wavespec}) integrated from 10 au, the largest scale probed in this study (see Methods). We find $C_{\rm n}^2 = 10^{-1.64\pm0.02}$ m$^{-20/3}$, slightly smaller than the value $C_{\rm n}^2 = 10^{-1.47\pm0.04}$ m$^{-20/3}$ found in a previous analysis of POEs between 2012 and 2019 \citep{2020ApJ...904...66L}. The ratio of the rms density fluctuations to the mean density is ${\rm rms}(n_e)/\bar{n}_e = 0.0041/0.12 = 0.034$.

\indent In the local ISM, within about 1~kpc probed by pulsar observations, the spectral coefficient for $\beta = 11/3$ is $C_{\rm n}^2 \approx 10^{-3.5}$ m$^{-20/3}$. \citep{1991Natur.354..121C, 2015ApJ...804...23K} The electron density of the local ISM has a scale height of about 1.6 kpc and mean density at mid-plane of 0.015 cm$^{-3}$ \citep{2020ApJ...897..124O}. Integrating the Kolmogorov wavenumber spectrum with $C_{\rm n}^2 = 10^{-3.5}$ m$^{-20/3}$ from a cutoff of 10 au yields ${\rm rms}(n_e) = 4.8\times10^{-4}$ cm$^{-3}$, which combined with $\bar{n}_e = 0.015$ cm$^{-3}$ gives  ${\rm rms}(n_e)/\bar{n}_e = 0.032$ for the local ISM, very close to our observed value for the VLISM. It has been suggested \citep{2019ApJ...887..116Z,Fraternale_2021} that the VLISM turbulence spectrum is a superposition of higher and lower amplitude Kolmogorov spectra of heliospheric and interstellar origin, respectively, which may explain why the spectral coefficient we infer is much larger than expected for the local ISM.

\indent The extremely weak, narrowband plasma wave emission reported here persists over about 10 au of interstellar space traversed by V1. The emission appears to be distinct from shock-generated plasma oscillations that were previously used to measure the local density of plasma outside the heliopause, and it may be generated by thermally or suprathermally excited plasma oscillations. The persistence of the emission through the most recent data published from V1 suggests that it may continue to be detectable. Continued tracking of the emission's frequency, bandwidth, and intensity will likely provide improved constraints on the emission's origin, and make it possible for V1 to quasi-continuously track the electron density distribution of the VLISM along its trajectory. This future work will improve our understanding of turbulence and large-scale structure in the VLISM.

\section{Methods}\label{sec:methods}

\subsection{Constructing the wideband spectra}
\indent The V1 PWS wideband receiver measures the voltage difference across the spacecraft antenna terminals at a rate of 28,800 samples s$^{-1}$ for about 48 s to construct a voltage time series that is stored for later transmission to ground. Voltage time series are measured about once a week, although over the course of the mission this time gap varies from about a day to a couple of weeks. The voltage time series are converted into a frequency-time spectrum by the Fourier transform method, which allows tuning of the output spectrum's frequency resolution. The resulting spectrum shows the relative electric field intensity for frequencies up to 10 kHz, the frequency roll-off imposed by the low-pass filter on the receiver. To construct the composite dynamic spectra shown in Fig.~\ref{fig:fullspec} and Fig.~\ref{fig:panelspec}, the spectra of the voltage time series were sorted into 3-day long bins and then averaged within each bin. The subsequent time and frequency resolution of the spectrum in Fig.~\ref{fig:fullspec} is 3 days and 0.011 kHz, respectively. A lower frequency resolution of 0.02 kHz was used to construct Fig.~\ref{fig:panelspec} in order to avoid contamination of the narrowband signal by spectral ringing.

\indent Due to automatic gain control (AGC), the noise baseline of the voltage time series measured by the wideband receiver varies between observations. To correct for this effect, we equilibrated the noise threshold of the spectra obtained from the voltage time series by subtracting the noise continuum from each spectrum. The 2.4 kHz supply interference line and its harmonics were masked prior to noise equilibration. Our spectra therefore do not represent the absolute electric field intensity and should only be interpreted as the intensity relative to the noise baseline.

\indent There are two main sources of data quality degradation in the V1 wideband spectra. The first is spectral interference that persists throughout the entire timespan since V1 entered the VLISM (and prior): this includes the 2.4 kHz supply interference line and its harmonics, and high-intensity, low-frequency noise that dominates the spectra up to about 1 kHz. We avoided this spectral interference by masking each spectrum at the corresponding frequencies. The second source is degraded telemetry performance that produces bit errors causing short, temporal  ($\lesssim 1$ s), wideband bursts that vary in intensity and can take up to all of the frequency band. For most of the interstellar mission, these wideband bursts usually occur just a couple of times per observation, are about 5 kHz wide, and are only up to an order of magnitude more intense than the baseline noise. However, at Voyager's great distance from the Sun, the SNR of the telemetry signal arriving at the NASA Deep Space Network (DSN) is increasingly close to the level at which it cannot be decoded. As a result, since 2018 there has been a dramatic increase in the number, frequency bandwidth, and intensity of these bursts, which has been temporarily mitigated by the addition of a fourth antenna to the DSN array. The subsequent decrease in overall data quality had a minimal effect on the 8-yr long composite spectrum shown in Fig.~\ref{fig:fullspec} due to its long timespan, and we further reduced the minor effects of these bursts by applying a 1D Gaussian smoothing kernel with $\sigma = 0.01$ kHz to the entire spectrum in Fig.~\ref{fig:fullspec}.

\indent In order to accurately detect the narrowband signal shown in Fig.~\ref{fig:panelspec}, we masked the wideband bursts in each 48 s long spectrum before binning and averaging all of the spectra to create the composite dynamic spectrum in Fig.~\ref{fig:panelspec}. We also ignored altogether individual spectra that are completely contaminated by interference bursts, which corresponds to spectra containing 10 or more bursts, or equivalently, a noise threshold that is more than 5 times the equilibrated noise threshold of the composite spectrum. Based on this noise threshold, zero observations were excised in 2017, 12\% of observations were excised in 2018, and 33\% were excised in 2019. The average amount of time between consecutive, 48 s long observations is 2.9 days. The spectrum must be averaged over at least one 48 s long observation in order detect the narrowband emission above a SNR of 2; hence 2.9 days is the shortest timescale on which the narrowband emission can be deemed continuous.
 
\subsection{Plasma density detection and analysis}
\indent A modified friends-of-friends (FoF) algorithm \citep{1982ApJ...257..423H} extracted the plasma frequency (and hence density) time series shown in Fig.~\ref{fig:timeseries} from the unsmoothed spectrum of the narrowband signal (see Fig.~\ref{fig:panelspec}b). The modified FoF algorithm searched for continuous lines in the spectrum by finding all adjacent local maxima within a 0.2 kHz-wide band around a given starting frequency. The line candidates returned by FoF were then sorted by average intensity, and the line with the highest average intensity corresponded to a line in frequency vs. time space that passed through all local maxima containing the new narrowband signal. Outlier maxima picked up by FoF were then excised by calculating a running mean of the line and removing all points more than 0.04 kHz away from the mean. The ``de-fluctuated" spectrum was calculated by shifting the spectrum of the narrowband signal so that the peak frequency of the signal extracted by FoF was aligned to a reference frequency of 3 kHz.

\indent A Gaussian process model was fit to the plasma density time series using the \texttt{GaussianProcessRegressor} module provided by \texttt{scikit-learn} \citep{scikit-learn}, with the prior set to a white noise kernel added to a Matern kernel. The Matern kernel takes the form:
\begin{equation}
    k(x_i,x_j) = \frac{1}{\Gamma(\nu)2^{\nu-1}}\Bigg(\frac{\sqrt{2\nu}}{l}d(x_i,x_j)\Bigg)^{\nu} K_\nu \Bigg(\frac{\sqrt{2\nu}}{l}d(x_i,x_j)\Bigg)
\end{equation}
where $K_\nu$ is a modified Bessel function, $\nu$ is set to 1 and $l$ is set to 10. The time series of mean, posterior values then gave a continuous temporal model for the frequency of the narrowband emission. In Fig.~\ref{fig:timeseries}, we show the Gaussian process model and the 95\% confidence intervals of the posterior. This Gaussian process model is fairly insensitive to the large data gaps in late 2019 and 2020, resulting in a smooth progression between the density values bordering these gaps. The model is also shown next to previously detected POEs in the bottom panel of Fig.~\ref{fig:fullspec}, but with the largest data gaps in the density time series indicated.

\indent The variance of electron density fluctuations is directly related to the density fluctuation power spectrum (Eq.~\ref{eq:wavespec}) integrated over wavenumber $q = 2\pi/L$, where the length scale $L$ is related to the spacecraft speed (3.6 AU/yr). For an outer scale $q_{\rm o}$ much less than the inner scale $q_{\rm i}$, the integrated 3D wavenumber spectrum has the form
\begin{equation}\label{eq:varne}
    {\rm Var}(n_e) = \frac{2(2\pi)^{4-\beta}C_{\rm n}^2 l_{\rm o}^{\beta-3}}{\beta - 3} \approx 3(2\pi)^{1/3}C_{\rm n}^2l_{\rm o}^{2/3},
\end{equation}
where $\beta>3$ and the second approximation is for $\beta = 11/3$. The spectral coefficient $C_{\rm n}^2$ for the VLISM was estimated using the observed variance of the density time series and $\beta = 11/3$. Since the largest length scale of the density time series is 10 au, we substitute 10 au for $l_{\rm o}$ in Eq.~\ref{eq:varne} when integrating over the spectrum. The error in $C_{\rm n}^2$ was estimated by simulating the observed density fluctuations as the sum of a slowly varying signal and normally distributed white noise, with the total variance of the simulated signal set equal to the observed density variance. The Var($n_e$) was then calculated for 1000 realizations of the simulated signal and used to calculate the fractional error on Var($n_e$) and $C_{\rm n}^2$. The error estimate for $C_{\rm n}^2$ thus represents a lower limit because it only accounts for the error in Var($n_e$) due to additive noise. Eq.~\ref{eq:varne} was also used to calculate the density fluctuation variance assuming $C_{\rm n}^2 = 10^{-3.5}$ for the local ISM.

\subsection{Data Availability}
The Voyager 1 data used in this work are archived through the NASA Planetary Data System (\url{ https://doi.org/10.17189/1519903}). Data and examples of the PWS data processing algorithms are also available through the University of Iowa Subnode of the PDS Planetary Plasma Interactions Node (\url{https://space.physics.uiowa.edu/voyager/data/}).

\section{Materials and Correspondence}
Correspondence and requests for materials should be directed to S.K.O.: sko36@cornell.edu

\section{Acknowledgements}
Authors S.K.O., J.M.C., S.C., and S.R.S. acknowledge support from the National Aeronautics and Space Administration (NASA 80NSSC20K0784). S.K.O., J.M.C., and S.C. also acknowledge support from the National Science Foundation (NSF AAG-1815242) and are members of the NANOGrav Physics Frontiers Center, which is supported by the NSF award PHY-1430284. The research at the University of Iowa was supported by NASA through Contract 1622510 with the Jet Propulsion Laboratory.

\section{Author Contributions}
S.K.O. conducted the data analysis and wrote the initial draft of the paper. J.M.C., S.C., S.R.S., and S.K.O. are NASA Outer Heliosphere Guest Investigators on the Voyager Interstellar Mission. D.A.G. is the Principal Investigator of the Voyager PWS investigation and W.S.K. is a Co-Investigator of Voyager PWS and is responsible for the initial processing of the data at the University of Iowa. All authors contributed to discussion of the results and comments on the draft.

\section{Competing Interests}
The authors declare no competing interests.

\end{document}